Anion height dependence of $T_c$ for the Fe-based superconductor

Y. Mizuguhci[1,2,3], Y. Hara[4], K. Deguchi[1,2,3], S. Tsuda[1,2], T. Yamaguchi[1,2], K. Takeda[5], H. Kotegawa[2,4], H. Tou[2,4] and Y. Takano[1,2,3]

1.National Institute for Materials Science, 1-2-1 Sengen, Tsukuba, 305-0047, Japan
2.Japan Science and Technology Agency-Transformative Research-Project on Iron-Pnictides (JST-TRIP), 1-2-1 Sengen, Tsukuba, 305-0047, Japan
3.Graduate School of Pure and Applied Sciences, University of Tsukuba, 1-1-1 Tennodai, Tsukuba, 305-8571, Japan
4. Department of Physics, Kobe University, 1-1 Rokkodai, Nada, Kobe, 657-8501, Japan
5. Department of Electrical and Electronic Engineering, Muroran Institute of Technology, 27-1 Mizumoto, Muroran, 050-8585, Japan

Abstract
We have established a plot of the anion height dependence of superconducting transition temperature $T_c$ for the typical Fe-based superconductors. The plot showed a symmetric curve with a peak around 1.38 Å. Both data at ambient pressure and under high pressure obeyed the unique curve. This plot will be one of the key strategies for both understanding the mechanism of Fe-based superconductivity and search for the new Fe-based superconductors with higher $T_c$.

1. Introduction

Since the discovery of superconductivity in $LaFeAsO_{1-x}F_x$, several types of Fe-based superconductors, basically composed of square lattices of $Fe^{2+}$, have been discovered.[1] They are categorized into following typical crystal structures in order of simple structure; FeSe-type (11-type), LiFeAs-type (111-type), $BaFe_2As_2$-type (122-type), LaFeAsO-type (1111-type) and $Sr_4Sc_2O_6Fe_2P_2$-type (42622-type).[2-5]

Among the research on Fe-based superconductivity, pressure study is one of the most interesting fields, because the superconducting properties of the Fe-based superconductors are much sensitive to applying pressure. The firstly discovered Fe-based superconductor $LaFeAsO_{1-x}F_x$ shows a positive pressure effect on the transition temperature $T_c$; the onset temperature $T_c^{onset}$ increases from 26 to 43 K under 3 GPa.[1,6] On the other hand, $NdFeAsO_{1-x}$, which has a structure analogous to LaFeAsO, shows a negative pressure effect on $T_c$; the $T_c$ suddenly decreases with applying pressure.[7] Fe-chalcogenide superconductor FeSe shows the most dramatic pressure effect among the Fe-based superconductors; the $T_c^{onset}$ increases from 13 to 37 K under 4-6 GPa.[8-11] The enhancement of $T_c$ in FeSe indicates that the increase of $T_c$ should be related to the change in the crystal structure, because the carrier density of the FeSe layer does not change with applying pressure due to its simple structure with stacking of only FeSe layers. Several parent phases of the Fe-based superconductors exhibit pressure-induced superconductivity. For 122-type $AFe_2As_2$ (A = Ca, Eu, Sr and Ba), pressure suppresses the antiferromagnetic ordering and induces superconductivity, and the pressure where the superconductivity is induced depends on a uniaxality of the applied pressure.[12-16] Fe-chalcogenide FeTe, which has a structure analogous to superconducting FeSe, is one of the parent phases of the Fe-based superconductor, and shows the antiferromagnetic ordering. Contrary to $AFe_2As_2$, superconductivity is not induced by applying nearly hydrostatic pressure up to 19 GPa for FeTe.[17,18] However, it was recently reported that the tensile-stressed FeTe thin film showed superconductivity at 13 K.[19] This fact strongly indicates the superconducting properties of the Fe-based superconductor are closely correlated with its crystal structure.

Here we focus on the anion height from the Fe layer as a probe to clarify the relationship between $T_c$ of the Fe-based superconductor and its crystal structure. This is motivated by the theoretical study that proposes the anion height from the Fe layer as a possible switch between high-$T_c$ nodeless and low-$T_c$ nodal pairings for the Fe-pnictide superconductors.[20] Based on this respect, the $T_c$ of the Fe-based superconductors should depend on the anion height. Therefore, we establish the plot of the anion height dependence of $T_c$ for the typical Fe-based superconductors, and discuss the relationship between the $T_c$ and the anion height.

2. Anion height dependence of $T_c$

We plot the anion height dependence of $T_c^{onset}$ for the typical Fe-based superconductors as shown in Fig. 1.[1,4,5,9,21-34] The data points exhibit a symmetric curve with a peak around 1.38 Å as indicated by the green hand-fitting line. The filled and open symbols in Fig. 1 indicate the data points at ambient pressure and high pressure, respectively. The plotted transition temperatures were basically estimated by the following two ways: either the temperature where the resistivity deviated from the linear extrapolation of resistivity above $T_c$, or the cross point of two lines as described in the inset of Fig. 3. Because of the difference in definition of $T_c^{onset}$ in the referenced papers, $T_c^{zero}$ was also displayed with the small light-blue circles. To establish a plot that could describe an intrinsic feature of the Fe-based superconductivity, we selected the typical Fe-based superconductors with a following policy; the valence of Fe should be close to 2+, and the $T_c$ is the highest in that system. In this respect, $KFe_2As_2$ was excluded from this plot because it is composed of $Fe^{1.5+}$.[35] We also excluded the superconductors whose Fe layer was substituted by the other elements, for example, Co-doped $BaFe_2As_2$,[36] because the Fe-site substitution should be unfavorable for higher $T_c$ due to randomness in the superconducting layer. In fact, as plotted in Fig. 1, the data points of Pt-doped $BaFe_2As_2$[33] and Co-doped $LaFeAsO$[34] deviated from the unique curve.

The anion height depends on the type of the superconducting Fe-anion layers, and it increases in order of FeP, FeAs, FeSe and FeTe. The FeP-based superconductors have the low anion heights, and show the lower $T_c$ compared to the FeAs-based superconductors.[5,21] Focused on 1111 system, the $T_c$ considerably increases from 7 to 26 K by the replacement of As for P with elevating anion height.[1,21] For the LaFeAsO system, the $T_c$ dramatically increases up to 55 K when the anion height increases and approaches to 1.38 Å with the La-site substitution by Nd or Sm, which has the smaller ionic radius than that of La.[23-26] After passing the summit, the $T_c$ decreases along the unique curve through the data points of $TbFeAsO_{0.7}$, $Ba_{0.6}K_{0.4}Fe_2As_2$, $NaFeAs$ and $LiFeAs$.[4,27,29,30] Also the data point of $FeSe_{0.57}Te_{0.43}$, which is almost optimally doped $FeSe_{1-x}Te_x$, obeyed the curve.[31] Therefore, the unique curve is applicable to 1111-, 122-, 111- and 11-type superconductors commonly. For the 42622-type superconductors, which have a thick blocking layer, the optimal $T_c$ has not been determined yet. However, the data points obey the unique curve, and seem to be located slightly outside of the curve,[5,28] suggesting the possibility of higher $T_c$ than 55 K. It might be due to the enhancement of two-dimensionality.

3. The cases under high pressure

This plot is applicable also to the pressurized system. The effect of the application of pressure is primarily to alter the physical structure, unlike the chemical doping process introduces changes in both the structure and total carrier density. In this respect, we can directly discuss the relationship between the structure and the $T_c$. In fact, the data under high pressure for NdFeAsO$_{0.85}$, which is the almost optimally doped 1111-type superconductor, obeys the unique curve. The $T_c$ decreases from 51 K (at ambient) to 41 K (at 5.6 GPa) according to the decrease of the anion height as indicated by orange open squares in Fig. 1.[37] Furthermore, the data points of both the pressurized SrFe$_2$As$_2$ and BaFe$_2$As$_2$, which show pressure-induced superconductivity, also correspond to the curve when the $T_c$ becomes highest under the optimal pressure.[38-40]

We discuss the case of FeSe, which shows dramatic pressure effect. The $T_c$ of FeSe at ambient pressure is lower than that of the optimally doped FeSe$_{1-x}$Te$_x$. This indicates that the conditions for superconductivity are not optimized in FeSe at ambient pressure. This respect is consistent with the data point located below the curve in Fig. 1. The pressure dependence of $T_c$ of FeSe shows an anomalous behavior as shown in Fig. 2. With increasing pressure, it shows an anomaly around 2 GPa, and then exhibits a big raise above 2 GPa.[11] The pressure dependence of Se height also exhibits an anomaly around 2 GPa.[9] We added these data points to Fig. 1. Interestingly, the data point approaches to the unique curve with increasing pressure, and corresponds to the curve above 2 GPa, at which the anomaly was observed in $T_c$ as shown in Fig. 2. In the respect that the data points above 2 GPa correspond to the unique curve, an intrinsic superconductivity might be induced by applying pressure above 2 GPa.

4. Pressure effects for FeTe$_{0.8}$S$_{0.2}$ and FeSe$_{0.53}$Te$_{0.47}$

We also investigated the pressure effect of FeTe$_{0.8}$S$_{0.2}$, which is located at the right side in Fig. 1. The polycrystalline sample of FeTe$_{0.8}$S$_{0.2}$ was synthesized using the solid-state reaction method as described in Refs. 32 and 39. The obtained FeTe$_{0.8}$S$_{0.2}$ sample was exposed to the air for several days, because the $T_c$ and the superconducting volume fraction of the polycrystalline FeTe$_{0.8}$S$_{0.2}$ sample was enhanced by the air exposure while the as-grown sample showed only a filamentary superconductivity, as reported in Ref. 41. By the air exposure, the $T_c^{onset}$ and $T_c^{zero}$ increased up to 9.2 and 6.6 K, respectively. Electrical resistivity under high pressure was measured by the four-terminal method using an indenter cell.[42] Daphne oil 7474 was used as a pressure-transmitting medium. The actual pressure was estimated from the $T_c$ of the Pb manometer.

Figure 3 shows the temperature dependence of resistivity for FeTe$_{0.8}$S$_{0.2}$ under high pressure up to 2.31 GPa. With applying pressure, the resistivity decreased and the temperature dependence of resistivity became metallic compared to that at ambient pressure. The $T_c^{onset}$ and $T_c^{zero}$ were estimated as indicated in Fig. 3 and plotted in Fig. 4 as a function of applied pressure. Both $T_c^{onset}$ and $T_c^{zero}$ decreased with increasing pressure, and the superconducting transition almost disappeared at 2.31 GPa. Since the data of FeTe$_{0.8}$S$_{0.2}$ is located on the gradual slope of the curve, we did not expect such a large pressure effect on $T_c$. It indicates that the data points of FeTe$_{0.8}$S$_{0.2}$ under high pressure do not obey the unique curve.

One of the differences between FeTe$_{0.8}$S$_{0.2}$ and FeSe is the existence/absence of a disorder of the anion site in the superconducting layer. The superconducting layer of FeSe is composed of only Se and Fe. On the other hand, FeTe$_{0.8}$S$_{0.2}$ contains two different anions of Te and S in its superconducting layer. Due to the difference of the ionic radius between Te and S, there should be the disorder of the anion height. In fact, the high-resolution x-ray single crystal diffraction for the FeSe$_{0.44}$Te$_{0.56}$ indicated the existence of significantly different anion heights of Te and Se with a differential $\Delta h_{Te-Se}$ = 0.24 Å.[43] As suggested above, the superconducting properties of Fe-based superconductor are much sensitive to the anion height. If the disorder of the anion height exists, the superconducting properties would be affected strongly by the disorder. In fact, the pressure effect of FeSe$_{0.57}$Te$_{0.43}$ does not obey the curve in Fig. 1; the $T_c$ increases from 14 to 23 K accompanying with the decrease of the anion height from 1.620 Å (at 0 GPa) to 1.598 Å (at 2 GPa). The data at 2 GPa deviates from the unique curve, indicating that the plot could not be applied to FeSe$_{0.57}$Te$_{0.43}$.[31] The concept which we proposed here could be applicable to the superconductor that does not contain the disorder at not only the Fe site but also the anion site.

5. Conclusion

We have established the plot of the anion height dependence of $T_c$ for the typical Fe-based superconductors. The plotted data points exhibited the unique curve with the peak around 1.38 GPa. The data under high pressure also corresponded to the curve for NdFeAsO$_{0.85}$, SrFe$_2$As$_2$, BaFe$_2$As$_2$ and FeSe, which do not contain the disorder of the anion height. On the other hand, the data under high pressure for both FeTe$_{0.8}$S$_{0.2}$ and FeSe$_{0.57}$Te$_{0.43}$, which contains disorder of the anion heights in its superconducting layer, did not obey the curve. This result suggests that the disorder of the anion height affects the superconducting properties for the Fe-based superconductors. To clarify that, the detailed research focused on the disorder of the anion site should be performed. The anion height dependence of $T_c$ will be one of the key strategies for both understanding the mechanism of Fe-based superconductivity and search for the new Fe-based superconductors with higher $T_c$.

Acknowledgement
    This work was partly supported by Grant-in-Aid for Scientific Research (KAKENHI).

Figure captions

FIG. 1. Anion height dependence of $T_c$ for the typical Fe-based superocnductors. Large symbols indicate the onset temperature. The zero-resistivity temperatures at ambient pressure are indicated by small light-blue circle. Filled diamonds indicate the data at ambient pressure. Open diamonds are the data of $SrFe_2As_2$ and $BaFe_2As_2$ under the optimal pressure. Open squares indicate the data of $NdFeAsO_{0.85}$ under high pressure (HP). The data of FeSe under HP are indicated by open circles. A filled circle indicates the data of $FeTe_{0.8}S_{0.2}$. Filled Green diamonds are the data points of Pt-doped $BaFe_2As_2$ and Co-doped LaFeAsO.

FIG. 2. Pressure dependence of both $T_c$ and anion height.

FIG. 3. Temperature dependence of electrical resistivity for $FeTe_{0.8}S_{0.2}$ under high pressure up to 2.31 GPa. The inset shows an enlargement of low temperatures. The definition of the $T_c$ is indicated in the inset.

FIG. 4. Pressure dependence of $T_c^{onset}$ and $T_c^{zero}$.

FIG. 1

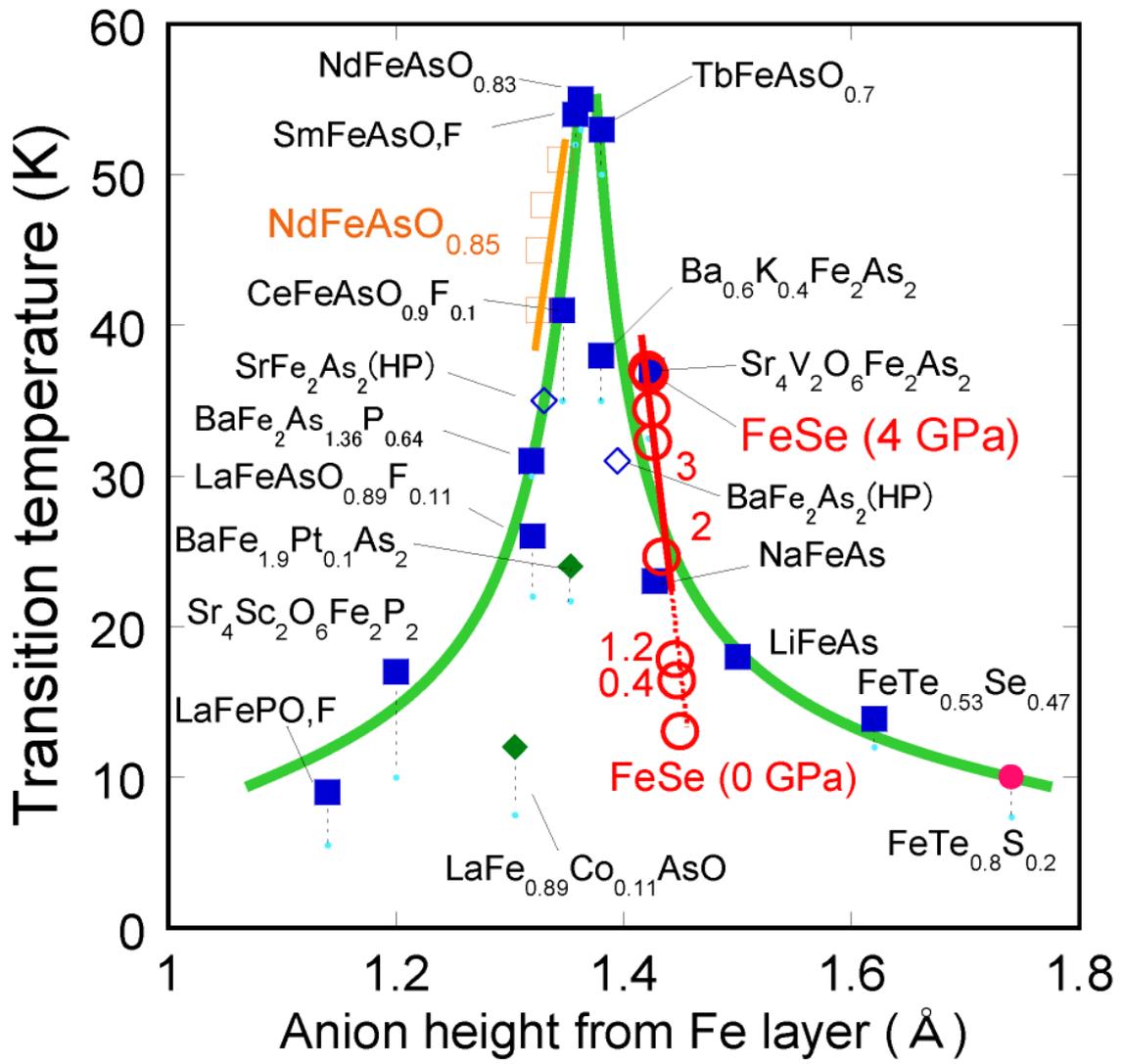

FIG. 2

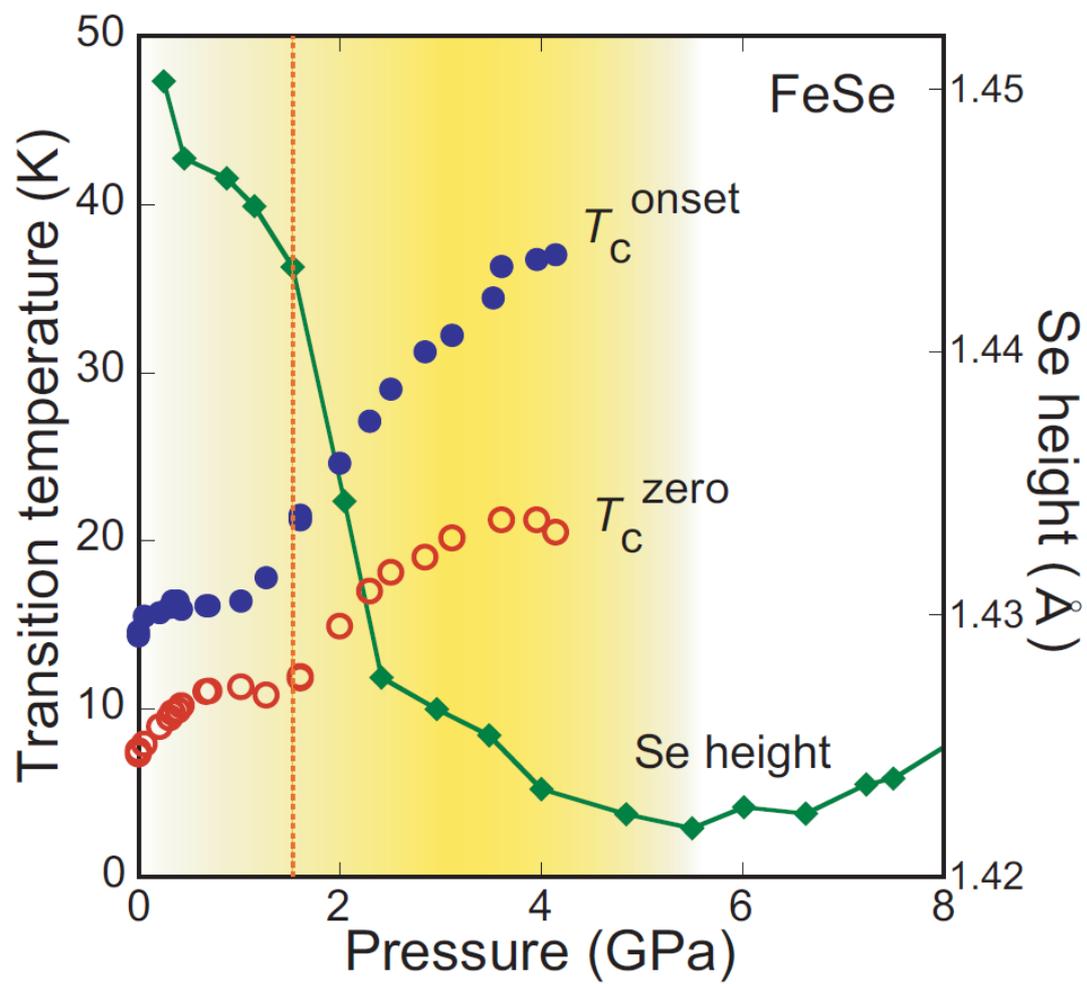

FIG. 3

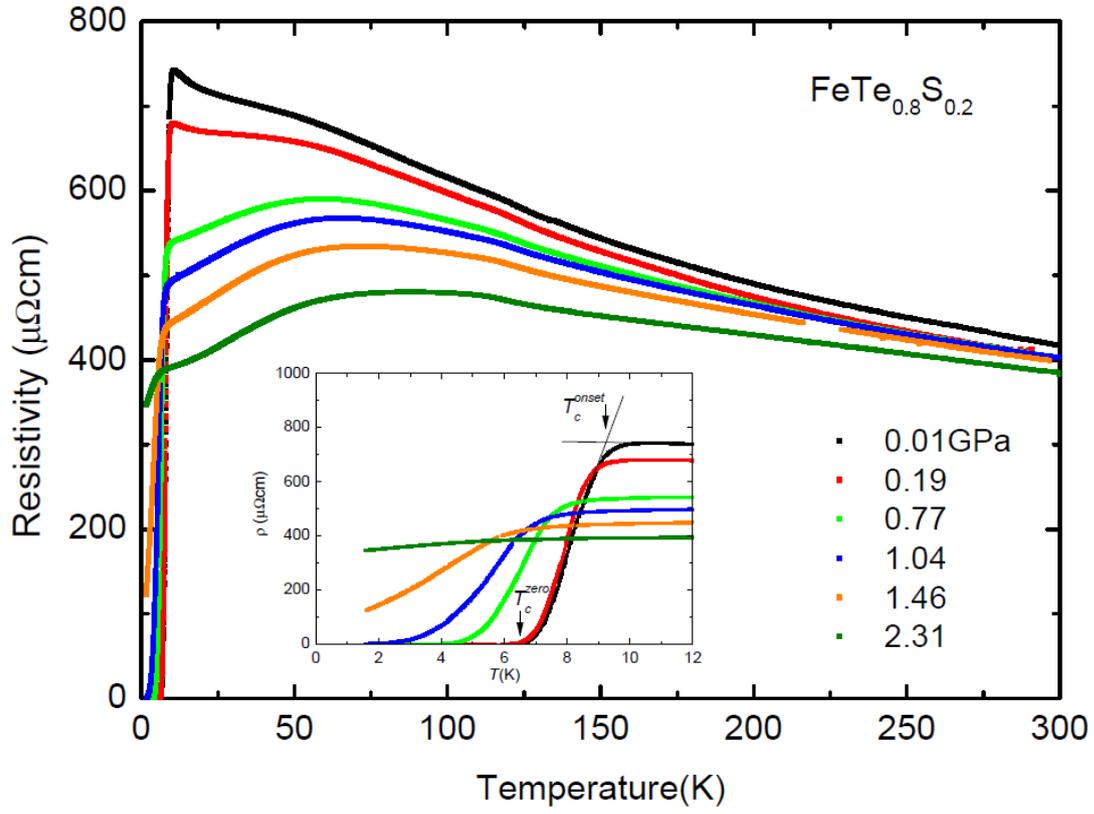

FIG. 4

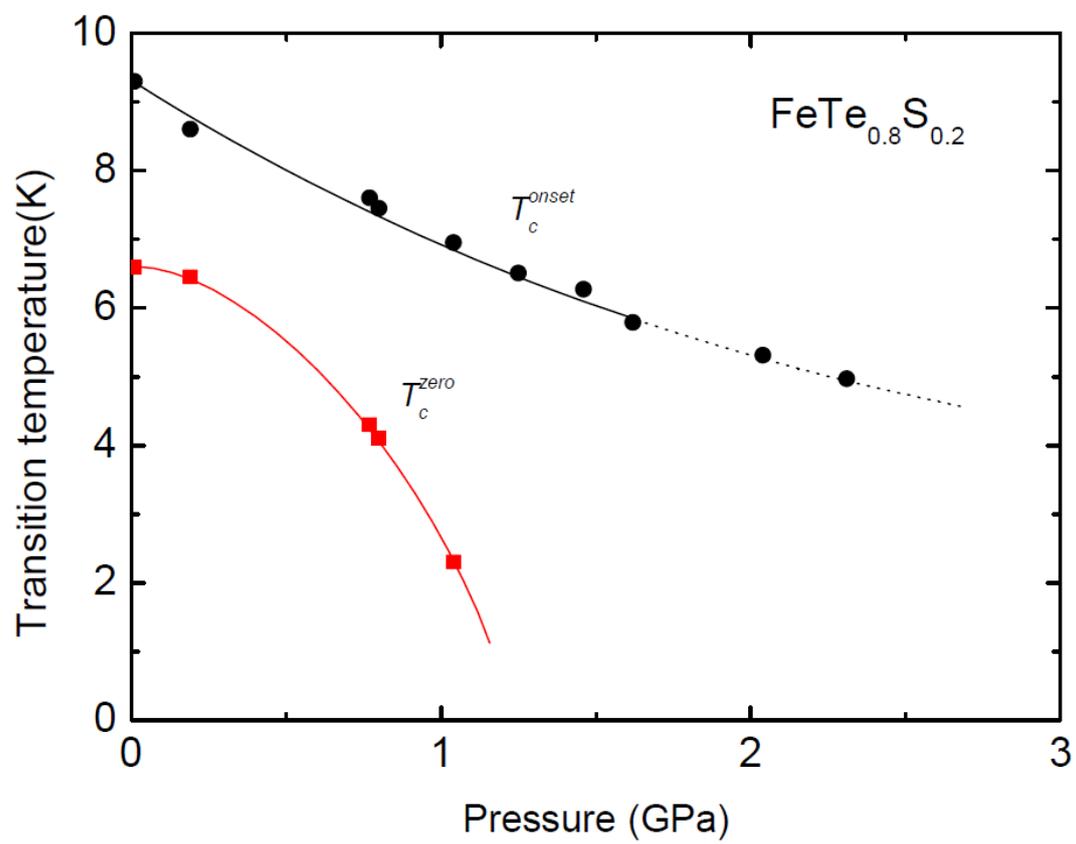